\documentclass[%
reprint,
showpacs,preprintnumbers,showkeys,
 amsmath,amssymb,
 aps,
pra,
]{revtex4-1}

\usepackage{graphicx}
\usepackage{dcolumn}
\usepackage{bm}

\begin{document}


\title{Unification of the conditional probability and semiclassical interpretations for the problem of time in quantum theory}

\author{J. C. Arce}
 \email{jularce@univalle.edu.co}
\affiliation{
Departamento de Qu\'imica, Universidad del Valle, A.A. 25360, Cali, Colombia
}%

\date{\today}

\begin{abstract}
We show that the time-dependent Schr\"odinger equation (TDSE) is the phenomenological dynamical law of evolution unraveled in the classical limit from a timeless formulation in terms of probability amplitudes conditioned by the values of suitably chosen internal clock variables, thereby unifying the conditional probability interpretation (CPI) and the semiclassical approach for the problem of time in quantum theory. Our formalism stems from an exact factorization of the Hamiltonian eigenfunction of the clock plus system composite, where the clock and system factors play the role of marginal and conditional probability amplitudes, respectively. Application of the Variation Principle leads to a pair of exact coupled pseudoeigenvalue equations for these amplitudes, whose solution requires an iterative self-consistent procedure. The equation for the conditional amplitude constitutes an effective ``equation of motion" for the quantum state of the system with respect to the clock variables. These coupled equations also provide a convenient framework for treating the back-reaction of the system on the clock at various levels of approximation. At the lowest level, when the WKB approximation for the marginal amplitude is appropriate, in the classical limit of the clock variables the TDSE for the system emerges as a matter of course from the conditional equation. In this connection, we provide a discussion of the characteristics required by physical systems to serve as good clocks. This development is seen to be advantageous over the original CPI and semiclassical approach since it maintains the essence of the conventional formalism of quantum mechanics, admits a transparent interpretation, avoids the use of the Born-Oppenheimer approximation, and resolves various objections raised about them.

\end{abstract}


\maketitle

\section{\label{sec:introduction} Introduction}

The time-dependent Schr\"odinger equation (TDSE) is commonly regarded as the fundamental equation of nonrelativistic quantum mechanics since it governs the spatiotemporal evolution of the wavefunction \cite{Bayfield99}. The time-independent Schr\"odinger equation (TISE), on the other hand, is viewed as a subsidiary of the TDSE because it can be derived from the latter by a separation of variables when the Hamiltonian operator is independent of time \cite{Bayfield99}.

However, there is something unsettling about time-dependent quantum mechanics: the spatial variables define a configuration space that contains the possible outcomes of position measurements performed on the system by an observer, which are given by the eigenvalues of the position operator, whereas the time variable plays the role of a parameter indicated by an \textit{external} clock (a \textit{classical} instrument) with no time measurements contemplated and, consequently, no time operator featuring in the formalism \cite{d'Espagnat99}. As a result of time not being an observable the time-energy uncertainty relation has a different character from the uncertainty relation involving any pair of noncommuting observables; in particular, the method for its derivation and its interpretation depend on the physical context. A historical account of the role of time in quantum mechanics is provided in Ref. \cite{Muga08} and the time-energy uncertainty relation is reviewed in Ref. \cite{Busch08}.

The involvement of an external clock implies that the TDSE really describes an \textit{open} system \cite{Breuer02}, albeit in the idealized limit that the back-reaction of the system on the clock is negligible [2(a)], so that relaxation and decoherence effects are imperceptible and the dynamics become unitary for practical purposes. Hence, it appears that if a truly closed system is to be considered its description should be fully based on the TISE \cite{Page83,Englert89,Briggs00}. If such an unorthodox view is taken seriously one is led to ask how time emerges from this fundamentally timeless framework. This question is particularly critical in quantum cosmology since, after all, the cosmos is the ultimate closed system.

In the field of quantum gravity the situation is further complicated because quantum mechanics and general relativity treat time in incompatible ways, the former as part of a fixed Newtonian background, the latter as part of a coordinatization of the spacetime manifold, which implies that it cannot be singled out in the theory since the equations of general relativity are covariant under spacetime coordinate transformations and physical results are independent of the coordinatization. The ``problem of time" in the context of quantum gravity is reviewed in Ref. \cite{Kuchar92}, but it must be realized that it is more general.

Several solutions to the problem of time have been proposed. One of them consists of coupling the system to a ``quantum clock" \cite{Peres80}, but it turns out that the ``time" defined by such clock exhibits some awkward characteristics like nonlocality, nonuniversality and quantization, and that a well-resolved time measurement must introduce a large perturbation into the system. The resulting state of affairs has been eloquently summarized by Peres [10(a)]: ``It thus seems that the Schr\"odinger wave function $\Psi(t)$, with its continuous time evolution given by $i\hbar\dot{\Psi}=H\Psi$, is an idealization rooted in classical theory. It is operationally ill defined (except in the limiting case of stationary states) and should probably give way to a more complicated dynamical formalism, perhaps one nonlocal in time. Thus, in retrospect, the Hamiltonian approach to quantum physics carries the seeds of its own demise."

The first sentence of Peres's statement is the observation that the TDSE is not a fully quantum-mechanical but rather a \textit{hybrid quantal-classical} equation of motion, because time appears to be a meaningful concept only at the classical level. Therefore, it seems natural to seek quantal variables that can play the role of clocks in the classical limit. This is the strategy followed by Briggs and coworkers, who have shown that the TDSE for the system results from the TISE of the clock plus system composite by first decoupling the clock variables from the rest of the degrees of freedom (DOFs), employing the Born-Oppenheimer (BO) approximation, and then using the WKB approximation for the clock variables \cite{Briggs00}. A somewhat similar method was used earlier by Englert \cite{Englert89}. Moreover, Briggs and Rost have shown that a time-energy uncertainty relation for the system can be deduced from the position-momentum uncertainty relation of the clock [8(a,e)]. Since a measurement is made at a particular instant (or interval) of time, in this approach the probabilistic interpretation (PI) \cite{d'Espagnat99} is introduced after the clock variables have been identified. Hence, it appears that the roles of the TDSE and the TISE have been inverted: the latter is the fundamental equation of nonrelativistic quantum mechanics, after all, since the former can be derived from it in the semiclassical limit. An analogous procedure can be applied in the relativistic quantum domain [8(a,b),9,11]; in particular, the semiclassical BO-WKB scheme has been widely employed in canonical quantum gravity after the pioneering work of Banks \cite{Banks85}, because its fundamental equation, the Wheeler-DeWitt equation \cite{DeWitt67}, does not involve a time variable explicitly.

More radical programs on the problem of time pursue the construction of a conceptually and formally consistent quantum theory that includes the PI but avoids any \textit{a priori} conception of time \cite{Page83,Englert89,Kuchar92,Barbour94,Dolby04,Gambini07,Corbin09}. Some of them seek an internal relational ``time", which in a sense resembles the external classical time, and an associated law that provides an effective description of dynamical evolution \cite{Page83,Englert89,Dolby04,Gambini07,Corbin09}, in tune with the second sentence of Peres's statement quoted above. A strategy of this sort is the conditional probability interpretation (CPI) of Page and Wootters \cite{Page83} whose basic tenet is to extract the dynamical evolution of the system's state from the conditional dependence of probabilities on the value of an internal clock variable.

All proposed solutions to the problem of time present their own conceptual and technical difficulties [6(c),9,17]. In this paper we point out that the BO-WKB and CPI schemes can be unified into a single framework, which is advantageous because it entails only a small reinterpretation of the conventional quantum-mechanical formalism, affords a derivation of the TDSE without recourse to the BO approximation, and resolves various objections raised on the original schemes. We work within the nonrelativistic formalism, but our development can be adapted to the quantum-gravitational realm.

In Sec. \ref{sec:statement}, the observation that the TDSE can be interpreted as an equation of motion conditioned by the classical state of a clock leads us to consider how the state of the system, given that the clock is in a particular pure state, can be represented. In Sec. \ref{sec:MCF} we achieve this rigorously by employing an \textit{exact} factorization of the composite system's eigenfunction in terms of marginal and conditional amplitudes \cite{Hunter75} for the clock variables and the remaining DOFs, respectively. Then, we employ the Variation Principle to generate exact coupled equations for such amplitudes \cite{Gidopoulos05}. In Sec. \ref{sec:clocks}, by introducing suitable clocks we reduce the conditional equation to the TDSE. Then, in this context we analyze the conditions required by physical systems to serve as good clocks. In Sec. \ref{sec:discussion} we critically discuss our development and its implications, in particular its advantages over the original BO-WKB and CPI schemes.

\section{\label{sec:statement} Statement of the Problem}

Let us consider a quantum system with spatial coordinates $\textbf{x}\equiv{x_1,x_2,\ldots}$. The TDSE
\begin{equation}\label{eq:TDSE}
\left(\hat{H}(\textbf{x},-i\hbar\nabla,t)-i\hbar\frac{\partial}{\partial t}\right)\Psi(\textbf{x}|t)=0
\end{equation}
dictates the evolution of its wavefunction with respect to the time $t$, which is a parameter indicated by an external classical clock. Although such clock ``ticks away" practically oblivious of the system, for the very concept of time to be meaningful there must exist at least a residual correlation between the quantal and classical states of the system and clock, respectively \cite{Page83,Englert89}, a situation that we may call ``classical correlation". Within the PI of quantum mechanics \cite{d'Espagnat99} such classical correlation can be construed as meaning that $\Psi(\textbf{x}|t)$  is the \textit{conditional} probability amplitude for the system to be in configuration $\textbf{x}$ \textit{provided that} the clock is in the configuration parametrized by $t$ \cite{Page83}, hence the bar we employ for separating the two kinds of variables. (For simplicity, when we speak of the probability of a system to be ``in a particular configuration" we actually mean ``in an infinitesimal neighborhood around a particular configuration".) Therefore, in light of these considerations we interpret the TDSE (1) as a conditional equation of motion.

Since the clock is constituted by atoms it must be ultimately governed by quantum laws as well. This elicits two closely related questions: First, how is it that the clock gets to behave classically so as to serve as a time-indicating device? This is the venerable problem of the classical limit of quantum mechanics [1,2(b)]. Second, how does the quantal correlation between the system and the clock become a classical correlation? In this paper we address the second question, taking for granted the classical limit. In particular, we set out to demonstrate that the TDSE is the dynamical law of evolution for the system that emerges in the classical limit from the conditional dependence of the configurational probability amplitude on the values of internal clock variables.

We begin by considering the system as a part of an essentially isolated composite quantum system, whose remaining part will eventually play the role of a clock. By ``essentially isolated" we mean that the composite system neither interacts nor is correlated in any way with the rest of the world \cite{Englert89}, i.e. it is a ``universe" in itself. Then, following other authors \cite{Page83,Englert89,Briggs00,Banks85,Barbour94,Dolby04,Gambini07,Corbin09,Unruh89} we postulate that the composite system is in a Hamiltonian eigenstate belonging to a Hilbert space $\mathcal{H}$, i.e.
\begin{equation}\label{eq:TISE}
\hat{H}\Phi(\textbf{x},\textbf{R})=E\Phi(\textbf{x},\textbf{R}),
\end{equation}
where $\textbf{R}\equiv{R_1,R_2,\ldots}$  are the spatial coordinates of the clock. Thus, here we depart from Peres's view expressed in the third sentence of his statement quoted in Sec. I. Next, we conveniently partition the composite Hamiltonian as
\begin{equation}\label{eq:Hamiltonian}
\hat{H}=\hat{H}_S(\hat{\textbf{x}},\hat{\textbf{p}})+\hat{H}_C(\hat{\textbf{R}},\hat{\textbf{P}})+\hat{H}_I(\hat{\textbf{x}},\hat{\textbf{R}}),
\end{equation}
where the first and second terms are Hamiltonians for the system and clock, respectively, and the third term represents the interaction between them. This interaction makes $\hat{H}$  nonseparable and, consequently, $\Phi(\textbf{x},\textbf{R})$  nonfactorizable into system and clock wavefunctions (pure states). In other words, since the system and the clock, considered independently, are open their states must be improper mixtures \cite{d'Espagnat99,Breuer02}. Therefore, the state of the system (clock) \textit{irrespective of the state of the clock (system)} can be represented by a reduced density matrix \cite{d'Espagnat99,Breuer02}. Nevertheless, to establish a connection with the conditional interpretation stated above, we must recognize that if the clock is to indicate a definite time it must be in a pure state. Then we pose the question: how can the state of the system be represented \textit{given that the clock is in a pure state}? Below we show that this can be done compactly by means of a suitably defined conditional probability amplitude.

Before proceeding, it is necessary to clarify our language. Our framework of timelessness and essential isolation precludes the \textit{a priori} incorporation of the conventional concept of measurement, simply because measurements are performed by observers at certain times, and forbids external observers, although internal ones could still be considered \textit{a posteriori}. That is why we adopt a ``realist" point of view and say that a system has a probability of \textit{being} in a certain configuration, instead of assuming an ``operational" standpoint and saying that a system has a probability of being \textit{observed} (measured) in a certain configuration [2(a)]. For classical systems the two viewpoints are practically equivalent, but for quantal ones they are not. Hence, we are led to renounce the Copenhagen interpretation \cite{d'Espagnat99} and subscribe to the relative-state (or many-worlds) interpretation of the wavefunction \cite{d'Espagnat99,Everett57}, as is commonly done in quantum gravity and cosmology \cite{Kuchar92,DeWitt67,Barbour94}.

\section{\label{sec:MCF} Marginal-Conditional Factorization of the Joint Eigenfunction}

We postulate that $|\Phi(\textbf{x},\textbf{R})|^2$  provides the joint probability density for the system \emph{and} the clock to be in configurations $\textbf{x}$ and $\textbf{R}$, respectively:
\begin{equation}\label{eq:PI}
|\Phi(\textbf{x},\textbf{R})|^2\equiv\rho_{joint}(\textbf{x},\textbf{R}).
\end{equation}
This is a well-defined quantity because $\textbf{x}$ and $\textbf{R}$ commute. Moreover, it is well known that the latter can be exactly factorized as [2(b),5]
\begin{equation}\label{eq:Bayes}
\rho_{joint}(\textbf{x},\textbf{R})=\rho_{mar}(\textbf{R})\rho_{con}(\textbf{x}|\textbf{R}),
\end{equation}
where
\begin{equation}\label{eq:defmar}
\rho_{mar}(\textbf{R}):=\int d\textbf{x}\rho_{joint}(\textbf{x},\textbf{R})
\end{equation}
is the marginal probability density for the clock to be in configuration $\textbf{R}$  \emph{irrespective of the configuration of the system}, and  $\rho_{con}(\textbf{x}|\textbf{R})$ is the conditional probability density for the system to be in configuration $\textbf{x}$  \emph{provided that} the clock is in configuration $\textbf{R}$ . (Throughout this paper, integrals such as that appearing in Eq. (\ref{eq:defmar}) are understood to be definite over the entire pertinent configuration space). The normalization of $\rho_{joint}(\textbf{x},\textbf{R})$,
\begin{equation}\label{eq:normrho}
\int d\textbf{R}\int d\textbf{x}\rho_{joint}(\textbf{x},\textbf{R})=1
\end{equation}
automatically implies the normalization of $\rho_{mar}(\textbf{R})$  and the \emph{local} normalization of $\rho_{con}(\textbf{x}|\textbf{R})$,
\begin{equation}\label{eq:normmar}
\int d\textbf{R}\rho_{mar}(\textbf{R})=1,
\end{equation}
\begin{equation}\label{eq:normcon}
\int d\textbf{x}\rho_{con}(\textbf{x}|\textbf{R})=1.
\end{equation}
By "local" we mean at a particular point of the $\textbf{R}$  configuration space.

Eq. (\ref{eq:Bayes}) can be extracted from the basic expression of the CPI \cite{Page83}, $Tr(\hat{P_A}\hat{P_B}\hat{\rho}\hat{P_B})=Tr(\hat{P_B}\hat{\rho}\hat{P_B})p(A|B)$, where $p(A|B)$  is the conditional probability for the system to have the value $A$ of the observable $\hat{A}$  given that the observable $\hat{B}$  has the value $B$, $\hat{\rho}$  is the density operator of the composite system and $\hat{P}_{A(B)}$  is the projection operator onto the $\hat{A}(\hat{B})$  subspace, by taking  $A=\textbf{x}$, $B=\textbf{R}$, $\hat{P}_{\textbf{x}}=|\textbf{x}\rangle \langle \textbf{x}|$, $\hat{P}_{\textbf{R}}=|\textbf{R}\rangle \langle \textbf{R}|$, and  $\hat{\rho}=|\Phi\rangle \langle \Phi|$. Clearly, $\hat{\rho}_{joint}=Tr(\hat{P_{\textbf{x}}}\hat{P_{\textbf{R}}}\hat{\rho}\hat{P_{\textbf{R}}})$  and  $\hat{\rho}_{mar}=Tr(\hat{P_{\textbf{R}}}\hat{\rho}\hat{P_{\textbf{R}}})$.

Following Hunter, we transfer the marginal-conditional factorization (MCF) of $\rho_{joint}(\textbf{x},\textbf{R})$  to the joint probability amplitude \cite{Hunter75}:
\begin{equation}\label{eq:MCF}
\Phi(\textbf{x},\textbf{R})=\text{X}(\textbf{R})\Psi(\textbf{x}|\textbf{R}).
\end{equation}
This factorization is exact as long as
\begin{eqnarray}\label{eq:defX}
\text{X}(\textbf{R})&:=& e^{i\alpha(\textbf{R})}\left(\int d\textbf{x}|\Phi(\textbf{x},\textbf{R})|^2\right)^{1/2}\nonumber\\
&\equiv& e^{i\alpha(\textbf{R})}\left\langle\Phi(\textbf{R})|\Phi(\textbf{R})\right\rangle^{1/2} ,
\end{eqnarray}
\begin{equation}\label{eq:defPsi}
\Psi(\textbf{x}|\textbf{R}):=e^{-i\alpha(\textbf{R})}
\frac{\Phi(\textbf{x},\textbf{R})}{\left\langle\Phi(\textbf{R})|\Phi(\textbf{R})\right\rangle^{1/2}},
\end{equation}
so that
\begin{equation}\label{eq:rhomar}
\rho_{mar}(\textbf{R})=|\text{X}(\textbf{R})|^2,
\end{equation}
\begin{equation}\label{eq:rhocon}
\rho_{con}(\textbf{x}|\textbf{R})=|\Psi(\textbf{x}|\textbf{R})|^2.
\end{equation}
(From Eq. (\ref{eq:defX}) onward we express integrals over the system configuration variables by means of angular brackets). The presence of the local phase $e^{i\alpha(\textbf{R})}$, with $\alpha(\textbf{R})$ real, in the definitions (\ref{eq:defX}) and (\ref{eq:defPsi}) permits that $\text{X}$  and $\Psi$  be complex even if $\Phi$ is real. The normalization conditions (\ref{eq:normrho})-(\ref{eq:normcon}) now read
\begin{equation}\label{eq:normPhi}
\int d\textbf{R}\left\langle \Phi(\textbf{R})|\Phi(\textbf{R})\right\rangle=1,
\end{equation}
\begin{equation}\label{eq:normX}
\int d\textbf{R}|\text{X}(\textbf{R})|^2=1,
\end{equation}
\begin{equation}\label{eq:normPsi}
\left\langle \Psi(\textbf{R})|\Psi(\textbf{R})\right\rangle=1.
\end{equation}

Evidently, $\text{X}(\textbf{R})=\left\langle \Psi(\textbf{R})|\Phi(\textbf{R})\right\rangle$ and $\Psi(\textbf{x}|\textbf{R})$ can be interpreted as marginal and conditional probability \emph{amplitudes}, respectively. It is quite remarkable that the joint eigenfunction of the \emph{entangled} system plus clock composite can be written in the product form (\ref{eq:MCF}), where the (nonlocal, EPR) quantum correlations \cite{d'Espagnat99,Page83,Englert89} of the system with the clock are embodied in the conditional-parametric dependence of $\Psi$ on $\textbf{R}$ . Hence, clearly $\Psi$ is not a wavefunction for the system in the conventional sense. To clarify the nature of this function let us first consider the mean value of an observable pertaining to the system alone:
\begin{eqnarray}\label{eq:expect}
\left\langle A_S\right\rangle&=& \int d\textbf{R}\left\langle \Phi(\textbf{R})|\hat{A}_S|\Phi(\textbf{R})\right\rangle\nonumber\\
&=&\int d\textbf{R}|\text{X}(\textbf{R})|^2\left\langle \Psi(\textbf{R})|\hat{A}_S|\Psi(\textbf{R})\right\rangle.
\end{eqnarray}
We observe that $\left\langle A_S\right\rangle$ is the average of the local quantities $\left\langle A_S(\textbf{R})\right\rangle\equiv\left\langle \Psi(\textbf{R})|\hat{A}_S|\Psi(\textbf{R})\right\rangle$  with the role of the distribution function played by $\rho_{mar}$ [Eq. (\ref{eq:rhomar})]. Since $\left\langle A_S(\textbf{R})\right\rangle$  can be interpreted as a conditional expectation value we see that $\Psi(\textbf{x}|\textbf{R})$ plays the role of a \emph{conditional wavefunction}, from which conditional quantities for the system can be evaluated following the prescriptions of conventional quantum mechanics \emph{locally}. Within the relative-state interpretation \cite{d'Espagnat99,Everett57} this function plays the role of a conditional relative state. Furthermore, taking into account Eq. (\ref{eq:normPsi}) we observe that $\int d\textbf{R}\left\langle \Psi(\textbf{R})|\Psi(\textbf{R})\right\rangle=\mathcal{V}$, with $\mathcal{V}$  the volume of the clock's configuration space. In Sec. \ref{sec:clocks} we will see that a clock can afford an infinite record of time only if $\mathcal{V}\rightarrow\infty$, so in that case $\Psi(\textbf{x}|\textbf{R})$ is not globally normalizable despite the fact that $\Phi(\textbf{x},\textbf{R})$ is. Therefore $\Phi$  inhabits the Hilbert space $\mathcal{H}=\mathcal{H_S}\otimes \mathcal{H_C}$, while, in general, $\Psi$ inhabits only the local Hilbert space $\mathcal{H_S}(\textbf{R})$. On the other hand, $\text{X}(\textbf{R})$ is a genuine wavefunction for the clock, inhabiting the Hilbert space $\mathcal{H_C}$, and we will call it the \emph{marginal wavefunction}.

The global normalization of $\Phi(\textbf{x},\textbf{R})$ [Eq. (\ref{eq:normPhi})] together with the local normalization of $\Psi(\textbf{x}|\textbf{R})$ [Eq. (\ref{eq:normPsi})] is all we need for a consistent PI of the formalism. Hence, our approach is free from the normalization problem afflicting the ``na\"ive" PI \cite{Kuchar92}.

It is worthwhile to mention the following point. For $\text{X}(\textbf{R})$ to have a zero at $\textbf{R}_0$ ~  $\Phi(\textbf{x},\textbf{R}_0)$ must vanish at all $\textbf{x}$ [see Eq. (\ref{eq:defX})]. Moreover, in this case, for $\Psi(\textbf{x}|\textbf{R}_0)$ to be well behaved $\Phi(\textbf{x},\textbf{R})$  must approach zero faster than $\text{X}(\textbf{R})$  as $\textbf{R}\rightarrow \textbf{R}_0$ at all $\textbf{x}$ [see Eq. (\ref{eq:MCF})]. This is an unlikely situation, so $\text{X}(\textbf{R})$ in general is a nodeless function. A more rigorous proof of this statement is given in Ref. \cite{Hunter81}.

We now apply the Variation Principle $\delta\langle\hat{H}\rangle=0$ to the derivation of the equations that govern the behavior of $\text{X}(\textbf{R})$ and $\Psi(\textbf{x}|\textbf{R})$ . Following Gidopoulos and Gross \cite{Gidopoulos05} we set up the functional
\begin{eqnarray}\label{eq:functional}
\mathcal{L}[\Phi]&\equiv&\int d\textbf{R}\left\langle \Phi(\textbf{R})|\hat{H}|\Phi(\textbf{R})\right\rangle\nonumber\\
&-&\int d\textbf{R}\lambda(\textbf{R})\left(\left\langle \Psi(\textbf{R})|\Psi(\textbf{R})\right\rangle-1\right)\nonumber\\
&-&\epsilon\left(\int d\textbf{R}|\text{X}(\textbf{R})|^2-1\right),
\end{eqnarray}
where the first term is the expectation value of the composite system's energy, the second term ensures the local normalization of $\Psi(\textbf{x}|\textbf{R})$  at every point of  $\textbf{R}$-space and the third term ensures the normalization of  $\text{X}(\textbf{R})$, with $\lambda(\textbf{R})$  and $\epsilon$ being local and global Lagrange multipliers, respectively. By imposing the extremization condition $\delta\mathcal{L}=0$ we obtain
\begin{eqnarray}\label{eq:mareq}
\bigg(&\hat{T}_C&+\sum_{n}\left\langle \Psi|m^{-1}_{n}\hat{P}_n|\Psi\right\rangle\hat{P}_n
+\left\langle \Psi|\hat{H}_S+\hat{H}_I+\hat{H}_C|\Psi\right\rangle\bigg)\nonumber\\
&\times&\text{X}(\textbf{R})=\epsilon\text{X}(\textbf{R}),
\end{eqnarray}
\begin{eqnarray}\label{eq:coneq}
\bigg(&\hat{H}_S&+\hat{H}_I+\hat{H}_C+\sum_{n}\left[\frac{m^{-1}_{n}\hat{P}_n\text{X}}{\text{X}} \right]\hat{P}_n
+\left[\frac{\hat{T}_C\text{X}}{\text{X}} \right]\bigg)\nonumber\\
&\times&\Psi(\textbf{x}|\textbf{R})=\frac{\lambda(\textbf{R})}{\rho_{mar}}\Psi(\textbf{x}|\textbf{R}),
\end{eqnarray}
with $m_n$ and $\hat{P}_n$ being the mass and momentum operator of the clock's $n$-th DOF, respectively, and $\hat{T}_C=\sum_{n}(2m_n)^{-1}\hat{P}^{2}_{n}$ the clock's kinetic-energy operator. In Eqs. (\ref{eq:mareq}) and (\ref{eq:coneq}) an operator within an angular-bracketed or square-bracketed expression, respectively, acts only on the function in front of it.

At the outset we make the following observations about these equations: First, they constitute a pair of \emph{exact} coupled pseudoeigenvalue equations, where the terms in parentheses on their left-hand sides play the roles of effective clock and system Hamiltonians, with $\text{X}$ and $\Psi$ being the respective eigenfunctions. (Note that $\text{X}$ and $\Psi$ are not eigenfunctions of Hamiltonians analogous to the nuclear and electronic ones appearing in the context of the BO approximation [8(a,b)].) Second, since the effective Hamiltonians are Hermitian the Lagrange multipliers $\epsilon$  and $\lambda(\textbf{R})$  are real. Third, together with Eqs. (\ref{eq:defX}) and (\ref{eq:defPsi}), they afford the local gauge freedom $\text{X}(\textbf{R})\rightarrow e^{i\gamma(\textbf{R})}\text{X}(\textbf{R})$, $\Psi(\textbf{x}|\textbf{R})\rightarrow e^{-i\gamma(\textbf{R})}\Psi(\textbf{x}|\textbf{R})$, with $\gamma(\textbf{R})$  real, leaving $\Phi(\textbf{x},\textbf{R})$ invariant. Finally, since  $\text{X}$  and $\Psi$ are unique up to a phase, for a given $\Phi$, each one of them possesses only one acceptable solution.

Let us now analyze the terms of these equations. In the marginal equation (\ref{eq:mareq}) the second and third terms of the effective Hamiltonian are \emph{mean-field} kinetic and potential couplings of the clock with the system, respectively. The vector-potential-like mean field $\left\langle \Psi|m^{-1}_{n}\hat{P}_n|\Psi\right\rangle$  vanishes if the phase carried by $\Psi(\textbf{x}|\textbf{R})$ is independent of $\textbf{R}$. We call ``effective clock potential" the quantity
\begin{eqnarray}\label{eq:defpot}
U_C(\textbf{R})&\equiv& V_C(\textbf{R})+\left\langle \Psi(\textbf{R})|\hat{H}_S+\hat{H}_I+\hat{T}_C|\Psi(\textbf{R})\right\rangle\nonumber\\
&=&\left\langle \Psi(\textbf{R})|\hat{H}|\Psi(\textbf{R})\right\rangle,
\end{eqnarray}
with $V_C(\textbf{R})$ being the potential present in $\hat{H}_C$ , which is seen to be analogous to a BO potential-energy surface. In the conditional equation (\ref{eq:coneq}) the fourth and fifth terms of the effective Hamiltonian are \emph{local} kinetic and potential couplings of the system with the clock, respectively.

In addition, let us left-multiply Eq. (\ref{eq:mareq}) by $\text{X}^*(\textbf{R})$  and then integrate over $\textbf{R}$ , taking into account Eq. (\ref{eq:normX}). With a little additional manipulation this yields $\epsilon=\int d\textbf{R}\left\langle \Phi(\textbf{R})|\hat{H}|\Phi(\textbf{R})\right\rangle=E$ . Next, let us left-multiply Eq. (\ref{eq:coneq}) by $\rho_{mar}(\textbf{R})\Psi^*(\textbf{x}|\textbf{R})$  and then integrate over $\textbf{x}$  and $\textbf{R}$, taking into account Eqs. (\ref{eq:normX}) and (\ref{eq:normPsi}). Comparison of both results indicates that $E=\int d\textbf{R}\lambda(\textbf{R})=\int d\textbf{R}\rho_{mar}(\lambda(\textbf{R})/\rho_{mar}(\textbf{R}))$ , so that $\lambda(\textbf{R})/\rho_{mar}$  can be interpreted as a local energy.

It is worthwhile to notice that if $\hat{H}_I$  vanishes identically, so that $\Psi$  becomes independent of $\textbf{R}$ , then Eq. (\ref{eq:mareq}) becomes the TISE for the isolated clock, with $E_C=\epsilon$, and Eq. (\ref{eq:coneq}), after multiplication by $\rho_{mar}$ followed by integration over $\textbf{R}$, becomes the TISE for the isolated system, with  $E_S=\int d\textbf{R}\lambda(\textbf{R})-E_C$.

\section{\label{sec:clocks} Clocks and the Unraveling of the TDSE}

Eq. (\ref{eq:coneq}) can be interpreted as an ``equation of motion" for the system's quantum state with respect to the clock variables consistent with the TISE (\ref{eq:TISE}). However, clearly this equation lacks \emph{evolution} since it does not involve a law that organizes the values of $\textbf{R}$  in a specific sequence. For the subsystem we have been calling ``the clock" to actually deserve that title it must provide such a law.

We call an ``ideal clock" a system that can indicate a definite time and provide an infinite record of it. Even if ideal clocks do not exist, the empirical fact that the TDSE provides a superb approximation for the description of nonrelativistic quantum phenomena in conventional laboratory settings \cite{Bayfield99} guarantees that at least good approximations to them do exist.

The present formalism does not involve the introduction of a time operator, which can be problematic \cite{Muga08,Busch08,Peres80,Unruh89}.  Instead, in Eq. (\ref{eq:coneq}) time must somehow emerge from the coupling of the system with the clock, which depends explicitly on the nature of the marginal wavefunction $\text{X}(\textbf{R})$.

The TDSE can be derived from Eqs. (\ref{eq:mareq}) and (\ref{eq:coneq}) following the same strategy as in the BO-WKB approach \cite{Briggs00,Kuchar92,Banks85}: first Eq. (\ref{eq:mareq}) is adiabatically decoupled from Eq. (\ref{eq:coneq}), then the WKB approximation for $\text{X}(\textbf{R})$ 
is used, and finally the classical limit for the clock variables in Eq. (\ref{eq:coneq}) is taken. Here we revisit this approach following a ``reverse-engineering" procedure, i.e., we assume the existence of ideal clocks and then determine the approximations required for the emergence of the TDSE from Eq. (\ref{eq:coneq}). Then we critically analyze the conditions under which typical physical systems can approach ideal clocks. This will turn out to be an instructive exercise.

Let us first consider a \emph{linear clock} [8(a),10(b)]. A simplified physical realization of such a clock is a structureless ball moving freely, say in the $Z$ direction, along a ruler with equally-spaced markings. The ``ticking" corresponds to the consecutive passing of the ball by the markings. The marginal eigenfunction then is the plane wave
\begin{equation}\label{eq:linclock}
\text{X}(\textbf{R})\rightarrow Ae^{\pm iP_{Z0}Z/\hbar},
\end{equation}
with $A$ being the constant amplitude and $P_{Z0}$ being the momentum eigenvalue of the ball. Substitution of Eq. (\ref{eq:linclock}) into Eq. (\ref{eq:coneq}) yields
\begin{eqnarray}\label{eq:coneq2}
\bigg(&\hat{H}_S&+\hat{H}_I+\hat{T}_C\pm\frac{P_{Z0}}{m}\hat{P}_Z+\frac{P^{2}_{Z0}}{2m}-\frac{\lambda(Z)}{|A|^2}
\bigg)\nonumber\\
&\times&\Psi(\textbf{x}|Z)=0,
\end{eqnarray}
where $m$ and $P^{2}_{Z0}/2m$  are the mass and kinetic energy of the ball, and $V_C(Z)$  was set equal to zero since the ball is free. In the classical limit  $P_{Z0}=h/\lambda\rightarrow\infty$, $P_{Z0}=m\dot{Z}$  and, integrating, $Z(t)=Z(0)+m^{-1}P_{Z0}t$, where the overdot represents the derivative with respect to time. By noticing that \cite{Briggs00}
\begin{equation}\label{eq:chain}
(\dot{Z}\hat{P}_Z)\Psi=-i\hbar\frac{dZ}{dt}\frac{\partial}{\partial Z}\Psi=-i\hbar\frac{\partial\Psi}{\partial t},
\end{equation}
we thus obtain
\begin{eqnarray}\label{eq:coneq3}
\bigg(&\hat{H}_S&+\hat{H}_I(\textbf{x},t)-\frac{\hbar^2}{2m}\frac{1}{\dot{Z}^2}\frac{\partial^2}{\partial t^2}\mp i\hbar\frac{\partial}{\partial t}\pm\frac{m\dot{Z}^2}{2}-\frac{\lambda(t)}{|A|^2}\bigg)\nonumber\\
&\times&\Psi(\textbf{x}|t)=0,
\end{eqnarray}
where the dependence on $Z(t)$  was equivalently expressed as a dependence on $t$. This is a differential equation of second order in the spatial and temporal variables for the quantum state of the system conditioned by the classical state of the clock, the latter being parametrized by $t$ and characterized by the constant momentum of the ball. The exactness of this equation depends only on the precision with which the physical clock approximates an ideal classical linear clock.

On taking the classical limit above time was introduced as a continuous parameter. Since the clock's ticking is discrete, this presupposes also taking the limit of the spacing between the markings going to zero. But nothing prevents us from working with a discrete time parameter [10(a,b)], which would turn Eq. (\ref{eq:coneq3}) into a differential-difference equation. Since our goal is to derive the (differential) TDSE, for now we will stick to a continuous time.

The classical $P_{Z0}=m\dot{Z}\rightarrow\infty$ limit allows three possibilities: (i) $m\rightarrow\infty,\dot{Z}\rightarrow\infty$, (ii) $m\ \text{finite},\dot{Z}\rightarrow\infty$, and (iii) $m\rightarrow\infty,\dot{Z}\ \text{finite}$. In all three cases the third term of Eq. (\ref{eq:coneq3}) vanishes (unless $\ddot{\Psi}\rightarrow\infty$, a possibility we do not entertain since we assume that $\Psi$ is well behaved). However, under typical laboratory conditions cases (i) and (ii) are not realistic because the speed of the classical ball is very small in comparison with the average speeds of the particles making up the quantal system. To analyze the more realistic case (iii) we first note that in the $m\rightarrow\infty,\dot{Z}\rightarrow 0$ limit the third term remains finite only if  $\ddot{\Psi}\propto\dot{Z}^\ell$, with $\ell\geq 2$. But $\ddot{\Psi}$ should depend smoothly on $\dot{Z}$, so if $\dot{Z}\agt 0$ then $\ddot{\Psi}\agt 0$ too. Therefore, in this case the time variations of $\Psi$ are very smooth.

After neglecting the third term of Eq. (\ref{eq:coneq3}), the last two terms can be eliminated by the gauge transformation \cite{Briggs00}
\begin{equation}\label{eq:gtrans}
\Psi(\textbf{x}|t)=e^{\frac{i}{\hbar}\left(\frac{m\dot{Z}^2t}{2}\mp\int^{t}\frac{\lambda(t')}{|A|^2}dt'\right)}\tilde{\Psi}(\textbf{x}|t),
\end{equation}
which finally produces the (first-order) TDSE for the system's gauged conditional wavefunction,
\begin{equation}\label{eq:gTDSE}
\left(\hat{H}_S+\hat{H}_I(\textbf{x},t)\mp i\hbar\frac{\partial}{\partial t}\right)\tilde{\Psi}(\textbf{x}|t)=0.
\end{equation}
Several comments about this equation are in order. First, it emerges independently of the eigenvalue $E$ in Eqs. (\ref{eq:TISE}) and (\ref{eq:mareq}); in particular, since the zero of energy is conventional, we could set $E=0$ in analogy with the Hamiltonian constraint of quantum gravity \cite{Kuchar92,DeWitt67}. Second, the sign of the time-derivative term is dictated by the sign of the exponent in Eq. (\ref{eq:linclock}), which is conventional as well. Thus, just like the distinction between forward and backward motion in Eq. (\ref{eq:linclock}) is meaningless, the distinction between past and future in Eq. (\ref{eq:gTDSE}) is also meaningless \cite{Barbour94}. Third, if $\hat{H}_I\approx 0$, i.e., if there is only a residual interaction between the system and the clock (all that is needed to establish the classical correlation alluded to in Sec. \ref{sec:statement}) the time-derivative term remains, as it originates from the ``chronogenic" fourth term in the left-hand side of Eq. (\ref{eq:coneq}), independently of the strength of the interaction. However, in this case the time-derivative term can be removed by the simple phase transformation $\tilde{\Psi}(\textbf{x}|t)=e^{\mp iE_St/\hbar}\psi(\textbf{x})$, which produces the TISE for the system $\hat{H}_S\psi(\textbf{x})=E_S\psi(\textbf{x})$ \cite{Bayfield99,Briggs00}.

Let us now examine the conditions under which our physical clock can approximate the ideal linear clock. Evidently the plane wave (\ref{eq:linclock}) can be a solution of Eq. (\ref{eq:mareq}) with good approximation only if $\left\langle \Psi|m^{-1}_{n}\hat{P}_n|\Psi\right\rangle\approx0$ and $U_C\approx\text{const}$ [see Eq. (\ref{eq:defpot})]. The first condition can be met only if the $\textbf{R}$ variations of $\Psi$ are very much smoother than the ones of X. This is indeed the case because the ball, being classical, possesses a comparatively very small de Broglie wavelength $\lambda$, whereas, according to the discussion above Eq. (\ref{eq:gtrans}), the  $\textbf{R}$ variations of $\Psi$, which dictate its time variations ($\dot{\Psi}\approx\text{const}$), are very smooth. This in turn implies that $\hat{T}_C\Psi\approx 0$, so that $U_C(\textbf{R})\approx\left\langle \Psi(\textbf{R})|\hat{H}_S+\hat{H}_I|\Psi(\textbf{R})\right\rangle$ which, for the reason just explained, can be considered as nearly constant as far as X($\textbf{R}$) is concerned. The neglect of the $\textbf{R}$ variations of $\Psi$ in Eq. (\ref{eq:mareq}) constitutes an \emph{adiabatic approximation}, which means that the back-reaction of the system on the clock is only potential, not kinetic. On the other hand, in Eq. (\ref{eq:coneq}) the $\textbf{R}$ variations of $\Psi$ were kept at least up to first order to get the TDSE. Hence, the system is kinetically affected by the clock. These asymmetrical influences are possible only if the energy of the clock ($m\dot{R}^2/2$) is very much larger than the one of the system \cite{Englert89,Briggs00,Banks85}, which in our case is consistent with the condition $m\rightarrow\infty,\dot{Z}\agt 0$ stated above.

The ideal-clock ansatz (\ref{eq:linclock}) violates the normalization conditions (\ref{eq:normPhi})-(\ref{eq:normPsi}). To make $\text{X}(Z)\in\mathcal{H}_C$ and recover these conditions the amplitude can be considered to be actually a slowly-varying envelope function, so that $dA/dZ\approx0$, $A(Z\rightarrow\pm\infty)\rightarrow0$, and Eq. (\ref{eq:coneq2}) is a good approximation. Then, Eq. (\ref{eq:linclock}) is just the WKB approximation \cite{Bayfield99,Briggs00,Kuchar92,Banks85}. Now, according to Eq. (\ref{eq:normPsi}) the normalization of the time-dependent wavefunction at each instant of time is guaranteed:
\begin{equation}\label{eq:normPsit}
\left\langle \Psi(\textbf{R})|\Psi(\textbf{R})\right\rangle\rightarrow\left\langle\Psi(t)|\Psi(t)\right\rangle=\left\langle\tilde{\Psi}(t)|\tilde{\Psi}(t)\right\rangle=1.
\end{equation}
The fact that the global integral $\int dt\left\langle\Psi(t)|\Psi(t)\right\rangle$ diverges, because $\mathcal{V}\rightarrow\infty$ for a linear clock, is immaterial for the consistency of this development [see the discussion following Eq. (\ref{eq:expect})].

The linear clock has the drawback that in practice the ball will eventually bounce from an obstacle, so this system can afford only a finite record of time. Let us explore a scattering-free model, the \emph{cyclic clock} [10(b)]. The archetypical physical realization of this clock is a structureless handle moving at constant angular speed over a circular framework, say in the $XY$ plane, with markings at regular angular intervals. In this case the marginal eigenfunction can be expressed as
\begin{equation}\label{eq:cyclock}
\text{X}(\textbf{R})\rightarrow Ae^{\pm iL_{Z0}\phi/\hbar},
\end{equation}
with $\phi$ and $L_{Z0}=m\hbar$ ($m=\pm0,\pm1,\pm2,\ldots$) being the azimuthal angle and perpendicular component of the angular momentum of the handle. Substitution of Eq. (\ref{eq:cyclock}) into Eq. (\ref{eq:coneq}) yields an equation similar to Eq. (\ref{eq:coneq2}), except that the fourth and fifth terms are replaced by $\pm I^{-1}L_{Z0}\hat{L}_Z\Psi(\textbf{x}|\phi)$ and $(L^{2}_{Z0}/2I)\Psi(\textbf{x}|\phi)$, with $I$ and $L^{2}_{Z0}/2I$ being the moment of inertia and rotational kinetic energy of the handle. Taking the classical limit $L_{Z0}=I\dot{\phi}$, $\phi(t)=\phi(0)+I^{-1}L_{Z0}t$, using the chain rule $(\dot{\phi}\hat{L}_Z)\Psi=-i\hbar(d\phi/dt)(\partial/\partial\phi)\Psi=-i\hbar\dot{\Psi}$, neglecting the third term, and performing the gauge transformation analogous to Eq. (\ref{eq:gtrans}) finally produces the TDSE (\ref{eq:gTDSE}) again. Since the global integral $\int d\phi\left\langle\Psi(\phi)|\Psi(\phi)\right\rangle=\mathcal{V}=2\pi$, now the time-dependent wavefunction is normalizable with respect to time. The conditions under which this physical system can approximate the ideal cyclic clock are completely analogous to the ones for the linear clock.

Nevertheless, this clock is able to indicate time uniquely only for $0\leq\phi\leq2\pi$ (mod $2\pi$). To extend its range of applicability we supplement it with a second handle (say the minutes one) independent from the first handle (say the seconds one):
\begin{equation}\label{eq:cyclock2}
\text{X}(\textbf{R})\rightarrow Ae^{iL_{10}\phi_1/\hbar}e^{iL_{20}\phi_2/\hbar}.
\end{equation}
For simplicity, we used only the positive sign in the exponent and will assume that the handles have the same moment of inertia. Substitution of this equation into Eq. (\ref{eq:coneq}) yields an equation analogous to Eq. (\ref{eq:coneq2}), whose fourth term is 
$I^{-1}(L_{10}\hat{L}_1+L_{20}\hat{L}_2)\Psi(\textbf{x}|\phi_1,\phi_2)$. By construction, in the classical limit the first handle performs an integer number of cycles $C_1$ within each cycle of the second handle, which requires $L_{10}=C_1L_{20}$. Using the chain rule the fourth term becomes
\begin{eqnarray}\label{eq:chain2}
(\dot{\phi_1}\hat{L}_1+\dot{\phi_2}\hat{L}_2)\Psi&=&-i\hbar\left(\frac{d\phi_1}{dt}\frac{\partial}{\partial\phi_1}+\frac{d\phi_2}{dt}\frac{\partial}{\partial\phi_2}\right)\Psi\nonumber\\
&=&-i\hbar\left(\frac{d\textbf{R}}{dt}\cdot\nabla\right)\Psi\nonumber\\
&=&-i\hbar\frac{\partial\Psi}{\partial t},
\end{eqnarray}
where $\textbf{R}=\phi_1\hat{e}_1+\phi_2\hat{e}_2$ and $\nabla=(\partial/\partial\phi_1)\hat{e}_1+(\partial/\partial\phi_2)\hat{e}_2$, with $\hat{e}_1$ and $\hat{e}_2$ being unit vectors. Note that had we taken the right-hand side of the first equality to be $-i\hbar(\partial/\partial t+\partial/\partial t)\Psi=-2i\hbar\dot{\Psi}$ we would have gotten a spurious factor of 2. The choice of Eq. (\ref{eq:chain2}) is the correct one since it considers the motion of the clock's state in configuration 2-space, instead of separate motions of each DOF in their respective 1-spaces, as it should because the time-dependent wavefunction is defined on the configuration space of the system, not on the physical space. Since the classical trajectory in configuration space returns to the same point after $C_1$ cycles of the fast handle, now this clock can indicate time uniquely for $0\leq\phi_1\leq C_12\pi$ (mod $C_12\pi$). Clearly, for a cyclic clock to be able to provide an infinite record of time it must possess an infinite number of handles. In this limit the clock's configuration space has an infinite volume and the time-dependent wavefunction is no longer normalizable with respect to time, $\int dt\left\langle\Psi(t)|\Psi(t)\right\rangle\rightarrow\infty$.

We can also consider any combination of linear and cyclic clocks, for example $\text{X}(Z,\phi)=A_Ze^{iP_{Z0}Z/\hbar}A_{\phi}e^{iL_{Z0}\phi/\hbar}$, which, according to the discussion above, is also able to provide a global time and give rise to the TDSE (\ref{eq:gTDSE}).

The last two examples convey two important messages: First, if a clock has $N$ DOFs its configuration $N$-space still defines a global time through the chronogenic fourth term in the left-hand side of Eq. (\ref{eq:coneq}). This is relevant because realistic instruments always interact with their environments, and since we are considering an essentially isolated composite system, we have to take into account such environment as part of the clock. This observation does not clash with the possibility of obtaining a \emph{multiple}-TDSE by successively shifting the quantal/classical boundary for selected variables [8(d)]. Second, a periodic system cannot provide an infinite record of time, since its trajectory in configuration space eventually returns to the same point, but a chaotic system can.

The linear and cyclic clocks studied above involve force-free motions (although in the latter there is a constraint). To see if a system involving a force can constitute a good clock as well, let us examine the \emph{harmonic clock} [8(c)], a simplified physical realization of which is a structureless ball moving in a one-dimensional parabolic potential. We first point out that to be able to synchronize this system with a force-free clock the spacing between markings along the physical trajectory of the ball can no longer be regular but must be adjusted according to its acceleration so that the ticks correspond to regular time intervals. In regions where the de Broglie wavelength of the ball is sufficiently short the marginal eigenfunction can be approximated by Eq. (\ref{eq:linclock}). Now the equation analogous to Eq. (\ref{eq:coneq3}) contains the additional term $(\hbar^2\ddot{Z}/2m\dot{Z}^3)\dot{\Psi}(t)$, but the TDSE (\ref{eq:gTDSE}) can still be derived from it following a procedure similar to the one employed for the linear clock [8(c)]. However, the WKB approximation breaks down around the classical turning points \cite{Bayfield99}, so it would seem that the TDSE emerged only away from such points. This breakdown is a strong objection raised on the semiclassical approach in the context of quantum gravity \cite{Kuchar92}. It seems to us that this objection arises from the assumption that time must be a continuous parameter. Nevertheless, if one considers a discrete time parameter, as discussed earlier, taking care to not place markings too near the turning points, this problem can be avoided. Afterwards, if desired, a continuous time parameter can be recovered by interpolation. Similar considerations can be made for systems with more complicated forces. At any rate, for the harmonic model Briggs and coworkers have elegantly shown that the turning-point issue can be circumvented by employing the coherent-states representation of the harmonic mode amplitude [8(c)]. Nevertheless, it must be noted that this system by itself cannot provide an infinite record of time either (be it discrete or continuous), due to the periodicity of the trajectory or the amplitude. In conclusion, the TDSE can emerge either from analog (continuous) or digital (discrete) clocks.

\section{\label{sec:discussion} Discussion}

We have accomplished our goal of demonstrating that the TDSE is the dynamical evolution law unraveled in the classical limit from a timeless formulation in terms of configurational probability amplitudes conditioned by the values of appropriately chosen internal clock variables. In so doing, we have unified the CPI \cite{Page83} and BO-WKB \cite{Briggs00,Banks85} schemes in a single framework.

The pair of basic Eqs. (\ref{eq:mareq}) and (\ref{eq:coneq}), first derived by Gidopoulos and Gross \cite{Gidopoulos05}, involve three main ingredients: (i) the postulate that the state of our essentially isolated composite system is governed by the TISE (\ref{eq:TISE}); (ii) the PI of the square modulus of the joint eigenfunction, Eq. (\ref{eq:PI}); and (iii) Hunter's MCF of the joint eigenfunction, Eq. (\ref{eq:MCF}) \cite{Hunter75}.

Ingredients (i) and (ii) are part of the original CPI, but (iii) is a departure from that approach, which is advantageous for the following reasons. First, the CPI is based on conditional \emph{probabilities} and the factorization (\ref{eq:Bayes}), entailing a modification of the standard formalism of time-dependent quantum mechanics, whereas our approach is based on conditional \emph{amplitudes}, leaving essentially intact such a formalism, although calling for a small reinterpretation thereof. Second, in the BO-WKB approach the ansatz resembling Eq. (\ref{eq:MCF}) is obtained from an expansion of the joint eigenfunction in an adiabatic basis and a subsequent decoupling of the system and clock wavefunctions, requiring the introduction of a largely unspecified average clock potential \cite{Briggs00}. To us such detour is unnecessary since Eq. (\ref{eq:MCF}) is exact. Furthermore, in the BO ansatz the system and clock have become disentangled, whereas in the MCF they still are fully entangled.

The subsequent unraveling of the TDSE from the exact conditional Eq. (\ref{eq:coneq}) involves three additional ingredients: (iv) the existence of good clocks [e.g. Eqs. (\ref{eq:linclock}), (\ref{eq:cyclock}) and (\ref{eq:cyclock2})], which is the basis of the so-called semiclassical approach in the field of quantum gravity \cite{Kuchar92,Banks85}, justified by the adiabatic approximation; (v) the classical limit of the clock variables; and (vi) the neglect of time derivatives of order higher than first [like the one appearing in Eq. (\ref{eq:coneq3})], justified in the limit of low rate of change of the clock variables. Ingredients (v) and (vi) are consistent with (iv) if the masses (or related quantities, e.g. moments of inertia) associated with the clock variables are very large.

Relaxation of (vi) entails dealing with a (Klein-Gordon-like) ``second-order TDSE" and all the interpretational problems it conveys \cite{Barbour93}. According to the discussion above Eq. (\ref{eq:gtrans}) these higher-order corrections should be very small in laboratory conditions, although analogous terms may become important in relativistic contexts \cite{Kuchar92}.

If the requirements necessary for the validity of the classical limit (v) are not fully satisfied it is not possible to obtain a well-defined time parameter through the chain rule of Eq. (\ref{eq:chain}). However, if the adiabatic approximation is still valid, Eq. (\ref{eq:coneq2}) (with or without $\hat{T}_C$) provides an effective ``equation of motion" with respect to the clock variables.

If the  $\textbf{R}$ variations of $\Psi$ do not happen to be very much smoother than the ones of X, the adiabatic approximation, with its associated neglect of the kinetic back-reaction of the system on the clock, breaks down and the WKB approximation (iv) is invalidated \cite{Kuchar92,Banks85}. To exactly account for this effect we have to fall back to the basic Eqs. (\ref{eq:mareq}) and (\ref{eq:coneq}), which contain all the system-clock couplings explicitly. In principle, self-consistent solutions of these equations can be constructed in an iterative fashion, analogously to the way the Hartree-Fock equations are solved in many-body theory. Within the original BO-WKB framework, the prospect of treating the system-clock couplings employing a Hartree-Fock-like self-consistent approach has already been pointed out by Anderson [11(c)]. Clearly, our treatment rigorously substantiates Anderson's notion.

Within refined versions of the CPI, the consequences of the clock's nonideality have been treated from a different perspective by Gambini and coworkers \cite{Gambini07}, who have shown that the equation of motion for the system's reduced density matrix contains a weak Lindblad-type term, causing the evolution to be not exactly unitary and, consequently, initial pure states to slowly decohere into mixed states. A direct calculation of the minimum decoherence has been performed by Corbin and Cornish \cite{Corbin09}. In our approach, either at the exact or approximate levels of Eqs. (\ref{eq:coneq}) or (\ref{eq:coneq2}), respectively, decoherence does not arise because the quantum state, $\text{X}(\textbf{R})$, of the clock is fully determinate.

It may be objected that the model clocks employed in Sec. \ref{sec:clocks} are too simplistic. After all, the quantum systems studied in laboratories are not ``classically correlated" with such simple devices but with the surrounding environment (which may include an actual clock). However, those examples teach us that any system whose configuration space possesses a subset of semiclassical variables ${\textbf{R}'}$, so that $\text{X}(\textbf{R})=\chi(\textbf{R}'')e^{\pm i\textbf{P}'_0\cdot\textbf{R}'/\hbar}$ , where $\chi(\textbf{R}'')$ is a slowly-varying function of the remaining DOFs, $[\hat{P}'_n,\hat{R}'_n]=i\hbar$, and $P'_0=\hbar/\lambda'\rightarrow\infty$, can serve as an approximate clock. Since any realistic environment is macroscopic ($N\rightarrow\infty$) and non-periodic, taking into account the discussion in the penultimate paragraph of Sec. \ref{sec:clocks} we conclude that the environment surrounding a quantum system provides the global time needed for the rise of the TDSE.

It has been pointed out that it is unclear how the \emph{a posteriori} PI of the BO-WKB approach is related to the \emph{a priori} PI of the underlying TISE (or Wheeler-DeWitt equation) \cite{Kuchar92}. Our formulation clarifies this issue because the PI of the square modulus of the time-dependent wavefunction emerges automatically from the combination of the PI of the square modulus of the joint eigenfunction [Eq. (\ref{eq:PI})], the MCF [Eq. (\ref{eq:MCF})] and the classical limit of the clock variables.

The validity of the original CPI has been forcefully questioned by Kucha\v{r}, on grounds that upon application of the condition the state gets knocked out of the Hilbert space, and, consequently, a ``frozen" propagator related to the question ``If one finds the particle at $\textbf{x}_0$ at the time $t_0$, what is the probability of finding it at $\textbf{x}_1$  at the time $t_1>t_0$?" is predicted [6(c),9(a)]. Refined versions of the CPI have been shown to be capable of answering correctly this question \cite{Dolby04,Gambini07}. In the present context Kucha\v{r}'s objection would mean that conditioning the clock to be in a particular configuration $\textbf{R}'$  would collapse its state to the distributional state $\delta(\textbf{R}-\textbf{R}')$. In our formulation this is not so, because such conditioning entails simply the evaluation at $\textbf{R}=\textbf{R}'$ of $\text{X}(\textbf{R})\in\mathcal{H}_C$,  $\Psi(\textbf{x}|\textbf{R})\in\mathcal{H}_S(\textbf{R})$, and  $\Phi(\textbf{x},\textbf{R})\in\mathcal{H}_S\otimes\mathcal{H}_C$. Furthermore, since our formulation is ultimately capable of yielding the TDSE, it naturally predicts the correct nonrelativistic propagator.

There is one final issue that we wish to emphasize. As mentioned earlier, the conditional Eq. (\ref{eq:coneq}) [or its approximate version Eq. (\ref{eq:coneq2})] lacks evolution. Thus, the system simply \emph{is} in all its possible conditional states; in Wootters's words, its history is ``condensed" [6(b)]. Then, by introducing the WKB approximation followed by the classical limit, a ``sleight of hand", time, and with it causality and history, magically appeared in Eq. (\ref{eq:coneq3}) \cite{Barbour93}. Thus, in a semiclassical approach the TDSE emerges \emph{phenomenologically} because the classical limit of the clock variables is taken for granted. A first-principles theory should explain how (classical) time emerges in a system from the internal spatial quantum correlations \cite{Page83,Englert89,DeWitt67}. That is to say that the classical limit of quantum mechanics should be sought from the TISE, not from the TDSE, which is already partly classical, as is commonly done [1,2(b)]. A hint as to what this theory may be like has been provided by Barbour \cite{Barbour94}.

\begin{acknowledgments}
The author is grateful to C.A. Arango for pointing out some typographical errors in the previous version.
\end{acknowledgments}

\bigskip


\end{document}